\documentclass[12pt]{article}
\usepackage[square,numbers]{natbib}
\usepackage{hyperref}
\usepackage{doi}
\usepackage{graphicx}  
\usepackage{adjustbox}
\usepackage{amsmath}
\usepackage{array}

\begin{document}
\begin{center}

{\bf Molecular dynamics simulation of melting of silicene}
\end{center}
\noindent
Tjun Kit Min$^a$, Tiem Leong Yoon$^{a,\ast}$, Thong Leng Lim$^b$

\vspace{2pc}
\noindent 
$^a$ School of Physics, Universiti Sains Malaysia, Penang, Malaysia \\
$^b$  Faculty of Engineering and Technology, Multimedia University, Melaka, Malaysia \\
\noindent 
$^\ast$ Corresponding author \\
\vspace{2pc}

\begin{abstract}
\noindent We report the melting temperature of free-standing silicene by carrying out molecular dynamics (MD) simulation experiments using optimized Stillinger-Weber (SW) potential by Zhang {\it et al.} \citep*{Zhang:PRB14}. The melting scenario of a free-standing silicene is well captured visually in our MD simulations. The data are systematically analyzed using a few qualitatively different indicators, including caloric curve, radial distribution function and a numerical indicator known as ‘global similarity index’. The optimized SW potential consistently yields a melting temperature of 1500~K for the simulated free-standing, infinite silicene.
\end{abstract}

\vspace{2pc}
\noindent{\bf Keywords}: Molecular dynamics, Melting, Silicene, Optimized Stillinger-Weber potential


\section{Introduction}
\noindent 
Silicene, a two-dimensional nanosheet made up of silicon atoms arranged in a honeycomb lattice, has been predicted theoretically by Takeda and Shiraishi \citep*{Takeda:PRB94}
in year 1994. Subsequent DFT calculations such as  \citep*{Guzman-Verri:PRB07}
and \citep*{Cahangirov:PRL09} revive the interest of the prediction by showing that silicene is indeed energetically stable, and a feasible possibility. Silicene, unlike grapehene which prefers sp2 hybridization, is not flat. Rather, due to the preference of the bounding for sp3 hybridization, the silicene sheet has a buckled configuration, where the out-of-plane buckle parameter is predicted to be {0.44 \AA} according to DFT calculations. Having a close resemblance to graphene, silicene offers many possibilities as a functional material of advanced applications, such as photovoltaic, and optoelectronic devices \citep*{Hu:PRB13},
thin-film solar cell absorbers beyond bulk Si \citep*{Huang:PRV13}
and hydrogen storage \citep*{Jose:PCCP11}.
One advantage of silicene over other 2D materials is that it is in principle easier to get integrated into nanodevices which are mainly silicon-based. 

Silicene is a relatively new 2D material, and was first  synthesized on supported substrates in a series of discovery since 2007 \citep*{Guzman-Verri:PRB07}.
Following the successful synthesis of silicene on supported substrate, many theoretical studies and simulations on the structural, mechanical, electronic and thermal properties of silicene on supported substrate have been published 
\citep*{LewYanVoon2016}. 

The structural properties of a free-standing silicene sheet get modified when grown on a substrate. The silicene experimentally synthesized so far are not free-standing but sitting on a substrate. We have yet to see any report of experimentally synthesized free-standing silicene. Having said that, investigation of free-standing silicene serves the purpose to understand the pristine system in the absence of interactions with surfaces. The understanding of the basic properties of silicene without the interference from surface interaction with substrate shall provide useful insight for higher level manipulation of silicene, such as the `van der Waals' heterostructures envisaged by \citep*{Geim:Nature13}. 


As far as we are aware of, very limited work on the melting behavior and thermal stability of free-standing silicene is reported in the literature. Bocchetti {\it et al.} simulated the melting behavior of free-standing silicene via Monte Carlo method with original and a modified version of Tersoff potential parameter set (known as ARK) for silicon atom 
\citep*{Bocchetti:JPCS14}. According to Bocchetti {\it et al.}, original Tersoff parameters for silicon atom results in a melting of the free-standing silicene at 3600 K, meanwhile the melting temperature obtained using ARK parameter set is $\sim$
1750 K. Berdiyorov {\it et al.} simulated the influence of defect on the thermal stability of free-standing silicene via MD using Reactive force-field (ReaxFF) 
\citep*{Berdiyorov:RSCA14}. He found that pristine silicene is stable up to 1500 K. As a general observation, melting properties and thermal stability of free-standing silicene obtained in MD simulations varies from cases to cases depending on the details of the simulation procedure. Most importantly, the simulation results are strongly forcefield-dependent. Apart from simulating thermal stability, MD simulation has been also been applied to investigate or predict thermal conductivity of both free-standing silicene. A wide range of potentials is employed in these simulations. For example, Zhang {\it et al.}  developed a set of Stillinger-Weber potential parameters specifically for a single-layer Si sheet to simulate the thermal conductivity 
\citep*{Zhang:PRB14}. Others use Tersoff potential with original parameter sets. 
In this paper we wish to contribute to the existing knowledge about the melting temperature of free-standing silicene sheet by carrying out MD simulations using modified Stillinger-Weber (SW) \citep*{Zhang:PRB14} 
potentials. Our MD simulations are carried out using the LAMMPS package
\citep*{Plimpton:JCP95}. To our best knowledge, optimized SW forcefield has never been used for simulating the melting scenario of an infinite silicene sheet. \\

\section{Methodology}
\noindent
To obtain an infinitely large free-standing silicene sheet, we first construct a bulk Si structure using a diamond unit cell with a lattice constant of 5.431~{\AA}. This is done by replicating a total of $30 \times 30 \times 5$ diamond unit cells to form a simulation box that is subjected to periodic boundary condition. The structure is then re-oriented such that the (1 1 1) surface is aligned along the $z$-axis. This is done because the Si diamond lattice in the (1 1 1) direction resembles a silicene sheet with a non-zero buckling parameter.
After re-orientation, all atom-containing $xy$-planes perpendicular to the $z$-axis are removed from the bulk until only a single sheet is left. A total of 72000 atoms is removed in the process. The removal of the silicon atoms creates a region of vacuum with 41.5~{\AA} thick along the $z$-direction on both sides of the remaining silicon atom plane, which is now resembling a free-standing silicene sheet with surface area of $130.56 \times 150.76$~{\AA}$^2$, consisting of 9600 atoms.
The sheet is infinite in extent (in the $x$- and $y$-directions) due to periodic boundary condition imposed in the MD simulation.

\begin{figure}[h]
\includegraphics[width=5.67in,height=2.38in]{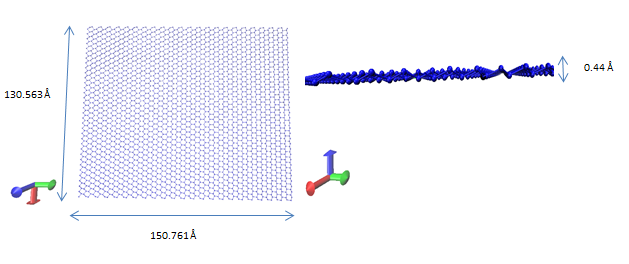}
\caption{Left: An infinitely large free-standing silicene sheet can be created by cutting out a silicon atom plane from the bulk Si diamond structure after re-orientating the (1 1 1) surface of the bulk into the $z$-axis. Right: The resultant structure has a 2D honeycomb shape with a dimension of 130.56~{\AA}$\times$150.76~{\AA} and buckling parameter of 0.44~{\AA}. The average bond length of Si-Si is 2.4~{\AA}.}
\label{image2}
\end{figure}

The data file of the free-standing silicene sheet is then inserted into the LAMMPS package to determine its melting point by carrying out simulated annealing simulations. SW potential is well known for its wide usage in semiconducting materials. It is first developed in 1985 for various phase of silicon crystal and it is in good agreement in predicting various physical properties regarding silicon crystals. The formalism of SW and its detail parameters is referred to the original Stillinger {\it et al.} paper \citep*{Stillinger:PRB85}. 
As it turns out, during our trial attempts to perform silicene melting simulation, conventional SW parameters of \citep*{Stillinger:PRB85}, as implemented in the LAMMPS codes, do not produce stable free-standing silicene sheet. On the other hand, using modified version of silicon SW potential by Zhang {\it et al.} \citep*{Zhang:PRB14} (the so-called optimized SW), we are able to obtain a stable 2D honeycomb structured silicene sheet upon energy-minimization at temperatures below its melting point. Moreover, the sheet turns out to have a buckling value of 0.44~{\AA} out of the $xy$-plane. The average bond length of Si-Si after energy minimization is 2.4~{\AA}. Hence, the optimized SW silicon potential by \citep*{Zhang:PRB14} is used exclusively in the present work. The details of the procedure are summarized as follow:
\begin{enumerate}

\item The free-standing silicene is relaxed via conjugate gradient minimization (energy minimization) technique. The resultant structure is depicted in Fig.~\ref{image2}.

\item Simulated annealing procedure is proceeded in NVT ensemble using the Nose-Hoover thermostat with a time step $\Delta t=0.5$~fs. The system is then equilibrated at 1 K for a period of $5\times 10^4 \Delta t$. Temperature is then raised to 300~K at a heating rate of $1 \times 10^8$~K/s. After reaching 300~K, the system is equilibrated for a period of $2.5 \times10^4 \Delta t$.

\item The stage is now set for raising the system's temperature to a large target temperature $T$ (one at which the silicene will surely melt). To raise the temperature from 300~K to $T$, we have to fix a heating rate. We have performed a convergence test by running trial MD melting simulations to obtain the melting temperature of the free-standing silicene at various heating rates by fixing the target temperature $T=2000$~K. From our convergence test, it is found that by choosing a heating rate of 1.7$\times10^{12}$ K/s, 
resultant melting temperature obtained is insensitive to variation in the heating rate pegged at the chosen value. The system is heated from 300~K for a total $2.0\times 10^6 \Delta t$ (equivalent to $1$~ns) at the mentioned heating rate until it reaches 2000~K. 


\item Evolution of the silicenes's MD  trajectory in the simulation is monitored visually as well as quantitatively. Radial distribution function, and caloric curve of the silicene sheet are numerically sampled and measured. In addition, we also measure a numerical descriptor, known as `global similarity index', $\xi _{i}$, to gauge the melting process. Detailed description of the global similarity index is discussed in the following.
\end{enumerate}

The definition of the global similarity index is based on generic chemical similarity idea for detecting configurational changes along the trajectory during the heating process. The evolution of the global similarity during the simulated annealing process, which carries statistical information of geometrical variations in the melting mechanism of the silicene will be elaborated in Section~\ref{results}. The functional form of $\xi_{i}$ is proposed to take the form of 
\begin{equation}\label{xi}
\xi_i={1 \over n}\sum_{s=1}^{n}(k_{s,i}+1)^{-1}
\end{equation}

\begin{equation}\label{ksi}
k_{s,i}=\lvert \sqrt{d_{s,i}} - \sqrt{d_{s,0}} \rvert 
\end{equation}
where $ d_{s,i}$ and $ d_{s,0}$ represent the sorted distance of atoms relative to the average positions (center of mass) of all the atoms in the system in the $i$-th (denoted as the subscript $i$) and the 0-th ``frame'' during a MD simulation, while $n$ corresponds to the number of atoms, which is an integer equals to the number of pairs of $k_{s,i}$. The value of $\xi_{i}=1$ corresponds to total identicalness between the configuration of the system in the $i$-frame as compared to that in the zeroth frame (which in practice is chosen to be a reference frame in which the system appears in a relatively perfect and stable configuration). $\xi_{i} \rightarrow 0$ infers vast difference.
 
Having defined the similarity index in  Eqs.~\eqref{xi},~\eqref{ksi}, one can optimize the sensitivity of the index with respect to degrees of freedom corresponding to few parameters such as translation and rotation of the particles relative to one another. Note that the average positions, a.k.a., mean, that enters the definition of the parameter $d_{s}$ can be non-uniquely defined. Different definitions of mean capture different aspects of configurational information contained in the system. Several definitions of mean are possible, namely, (a) arithmetic mean [see Eq.~\eqref{am}], (b) harmonic mean [see Eq.~\eqref{hm}] and (c) quadratic mean [see Eq.~\eqref{qm}]. 

\begin{equation}\label{am}
\bar x = \frac{x_1 + x_2 + \cdots + x_n}{n}
\end{equation}

\begin{equation}\label{hm}
\bar x = \frac{1}{
{1 \over x_1 }+ {1 \over x_2 } + \cdots + {1 \over x_n }
}
\end{equation}

\begin{equation}\label{qm}
{\bar x} = \sqrt{\frac{x_1^2 + x_2^2 + \cdots + x_n^2}{n}}
\end{equation}
However, during a melting process, variations in the configuration of the atoms \textit{a priori} are not known. In principle, variation in a certain mode of motion among the atoms could be more sensitively picked up by a particular definition of mean then the other. To cover all possibilities, the parameter $d_{s}$ that enters the definition of $\xi_{i} $ in Eqs.~\eqref{xi},~\eqref{ksi} is calculated by averaging overall of the above three types of mean, i.e., 
\begin{equation}
{\bar \xi_{i}}=\frac{1}{N_{\mathrm{COR}}} \sum_{\mathrm{COR}}^{} \xi _{i}^{\mathrm{COR}},
\end{equation}
where $\mathrm{COR}$ stands for center of reference, $\mathrm{COR}$=$\{$arithmetic mean, harmonic mean, quadratic mean$\}$, $ N_{\mathrm{COR}}=3$, $\xi_{i}^{\mathrm{COR}}$ is global similarity index defined based on a $\mathrm{COR}$ mean. Including all types of $\mathrm{COR}$ can in principle increase the sensitivity as well as accuracy of the index in capturing variations in the geometrical configuration of the system. Similarity index of the silicene frames in the MD simulation during its heating process is captured in successive temporal sequence. If the silicene sheet remains intact, its geometrical configuration should be very close to the one at the initial frame, i.e., ${\bar \xi_{i}} \approx 1$. If the silicene melts, we would expect ${\bar \xi_{i}}$ to deviate from 1 and in the extreme case, becomes 0. In the following we shall drop the bar symbol in the definition of $\xi_{i}$ when the index is referred. As it turns out, chemical similarity is rarely used for very large system \citep*{Baldi:JCIM10}
In the present case, the system being simulated consists of 9600 Si particles. This is considered a moderately large one. Global similarity index (which is a modified form of chemical similarity index) as defined above has a high sensitivity to detect even minor distortions in the geometrical configuration of silicene, and hence a very convenient tool to pinpoint the location of melting point during the temporal evolution of the MD. 

It is commented that caloric curve and the other two quantities, $g(r), \xi_i$, are two qualitatively different indicators for gauging the melting process. The former is a thermodynamical quantity whereas $g(r), \xi_i$ embed only geometrical information of how atoms distribute in 3D space. 

\section{Results and discussion}\label{results}

\begin{center}\begin{figure}[p]
\begin{center}
\includegraphics[width=4.6in,height=3.5in]{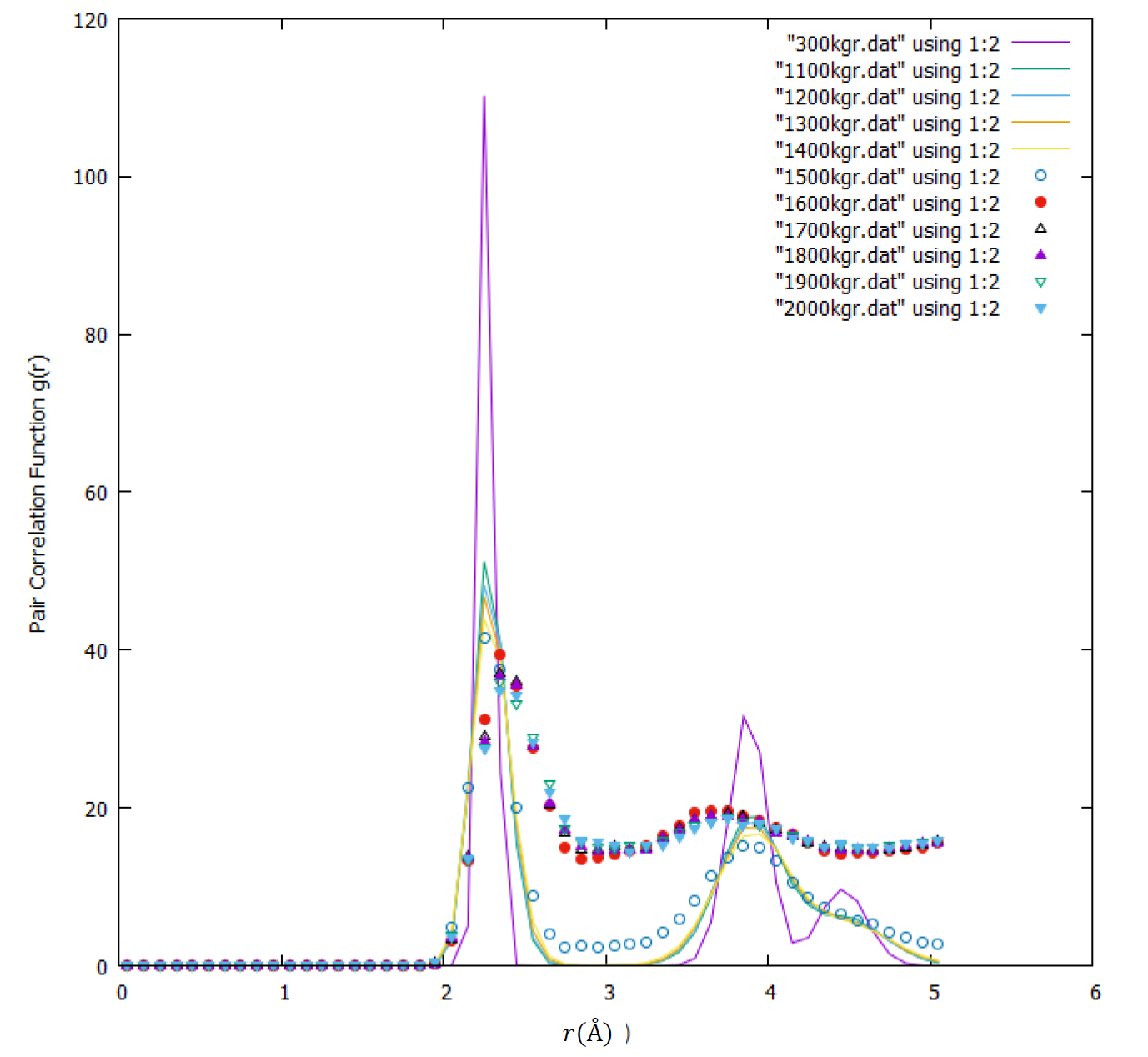}
\end{center}
\caption{Pair correlation function of the system at various temperatures. 
Two groups of curves are clearly visible. The group comprises of 300, 1100, 1200, 1300 and 1400 K curves (continuous, colored lines) is qualitatively distinctive from the group comprises of 1500 - 2000 K curves (filled and open circles and triangles). 
}
\label{figPCF}
\end{figure}
\end{center}

We first present the results of pair-correlation function, $g(r)$, measured in the simulation. Each pair-correlation function shown in Fig.~\ref{figPCF} is obtained by sampling the data in the MD simulation at that specific temperature. At room temperature, a sharp peak is measured at $\sim$2.5~{\AA}, indicating an average bond length of silicene at this value. As temperature increases, the peak at 2.5~{\AA} is lowered and begins to widen. 

The $g(r)$ curves with $T\leq 1400$~K are clearly distinctive from those with $T\geq 1500$~K. The former has their first peaks located at 2.5~{\AA}, while subsequent peaks are also clearly visible. In addition, $g(r)$ is  approximately zero between the first and second peaks. This group of curves essentially share a qualitatively similar geometrical distribution. We hence conclude that silicene sheet with temperature $T\leq 1400$~K remains intact and has not shown signs of complete melting. 

However, for curves with $T\geq 1500$~K, the first peak shifted to a larger position at $\sim$2.8~{\AA}, and the secondary peaks disappear. In addition, $g(r)$ beyond the first peak levitated to non-zero values, and remains relatively flat and constant beyond $\sim 3.0$~{\AA}. The disparity between these two groups of curves indicates that a qualitative change in the geometrical distribution of the silicene sheet has occurred when temperature crosses the 1400-1500~K benchmark. At 1500~K melting process has taken place. The features of the $g(r)$ curves for $T\geq 1500$~K are indicative evidence that the silicene sheet has melted into a liquid state at $T\geq$1500~K. The liquid state of the melted silicon system beyond 1500~K can be visualised in the last few snaphots in Fig.~\ref{t2}. These are liquid blobs made up of silicon atoms sticking together with an average bond length of $\sim$ 2.8 {\AA}. 

Figures shown in Figs.~\ref{t1} and \ref{t2} visually depict progressive melting process of the silicene sheet in the simulated annealing simulation obtained when the system is heated from 300~K towards 2000~K. At room temperature, the silicene sheet is thermally stable and maintains its hexagonal shape at 300~K. It is not perfectly flat though, with some puckering distortion along the direction perpendicular to the surface if viewed from the side. The equilibrium bond length of Si-Si is 2.5~{\AA}. At room temperature, the Si particles vibrate at their respective equilibrium positions, maintaining a defect-free silicene sheet. As the temperature increases, the particles begin to vibrate more violently and vigorously around $\sim$1100~K. Visually, it can be observed that vacancies begin to appear at selected areas on the silicene sheet as Si particles are ejected from their existing sites due to the thermal energy gained from the reservoir. At this temperature, however, the structure remains overall intact. 
For temperature at or above 1500~K, disintegration of the silicene sheet is clearly sighted. For temperatures less than the melting temperature, no sighting of silicene melting is observed. In other words, visual sighting provides another direct indication of melting at and above $T=1500$~K. 


Caloric curves sampled for the simulated silicene sheet at various temperatures provide an independent measurement of the melting point. It is seen that in 
Fig.~\ref{fig2} an abrupt drop in the potential energy plot occurs at $\sim$1500~K. From here onward, the sheet begins to tear apart and lose molecular cohesion, signaling the onset of full-fledge melting process. Concurrent to the abrupt drop in the caloric curve, disintegration of the sheet at $\sim$1500~K is also visually sighted in MD simulation video. The atoms of the silicene sheet break apart, forming disconnected 3D amorphous blobs, such as that displayed in the last few snapshots in Fig.~\ref{t2}. 


The melting process is now analyzed in terms of global similarity index, which is yet another independent measurement of the melting temperature. The structure during heating process is frame-by-frame compared with the reference structure at 300~K (the `0-th frame').
The curve of global similarity index as a function of temperature is shown in Fig.~\ref{figGSI}. As temperature increases, the index gradually decreases. At 1500~K the global similarity index drastically drops from $\sim 1$ to 0.02, inferring that the configuration of the system at or above 1500~K is drastically different from that at the reference configuration at 300~K. The melting point as indicated by global similarity index coincides with that obtained from other gauging indicators discussed above.

At this point we wish to comment that the emergence of local point vacancies, visually sighted at $\sim1100$~K, when silicon atoms leaving their existing sites, cannot be effectively captured by all melting indicators.

The final structure obtained when the system is quenched to 300~K is as shown in Fig.~\ref{fig5}. It is found that the resultant structure does not revert to its original planar form. Instead, these atoms condensed into some kind of 3-D amorphous ``columns''.

We have attempted to run another round of independent MD simulation on the melting of silicene sheet, with all the simulation specifications being exactly the same as that reported but with the NVT ensemble replaced by the NPT ensemble. It is reckoned that the implementation of NPT ensemble in a MD simulation effectively mimics an ‘adjustable periodic boundary’. The volume of the simulation in the NPT ensemble is monitored as a function of temperature. The variation in the volume as the temperature increases is as depicted in Fig.~\ref{S1}. It is found that the volume of the silicene fluctuates along the course where the temperature is annealed from below toward the melting temperature. Once it reaches the melting temperature the volume displays an abrupt drop, resulting in an apparent decrease in the volume of the silicene simulation. This observation is consistent with that found in bulk silicon melting process \cite{Endo:HTHP_03}. It is found that the melting temperature as measured in the NPT and NVT ensemble remains consistently similar. The only slight observational difference of the simulation in both ensembles is that the silicene in the NPT appears to display an undulation of a relatively larger amplitude before the melting point. In both ensembles, the top view of silicene reveals that the sheet remains intact in its intrinsic hexagonal honeycomb shape before melting.

\section{Conclusions}
We have performed a MD experiment to measure the melting point of a finite but large free-standing silicene sheet using optimized SW potential by Zhang {\it et al.}. The melting scenario of the silicene sheet has been visually captured. Analysis of the statistical data sampled from the MD simulation, including caloric curve, radial distribution function and global similarity index, reveals that the silicene sheet melts at 1500 K. This result serves the purpose of being an independent measurement of the melting temperature of free-standing silicene. Our value is consistent with that obtained by Berdiyorov {\it et al} who uses ReaxFF \citep*{Berdiyorov:RSCA14}, but is in contrast to \citep{Bocchetti:JPCS14} which predicts 1750~K using ARK parameter set. The prediction of free-standing silicent sheet will serve as a reference information to be verified by experiments when they become available.

\listoffigures
\bibliographystyle{unsrtnat}




\begin{center}
\begin{figure}[h]
\begin{center}
\includegraphics[width=9.6cm,height=14.4cm]{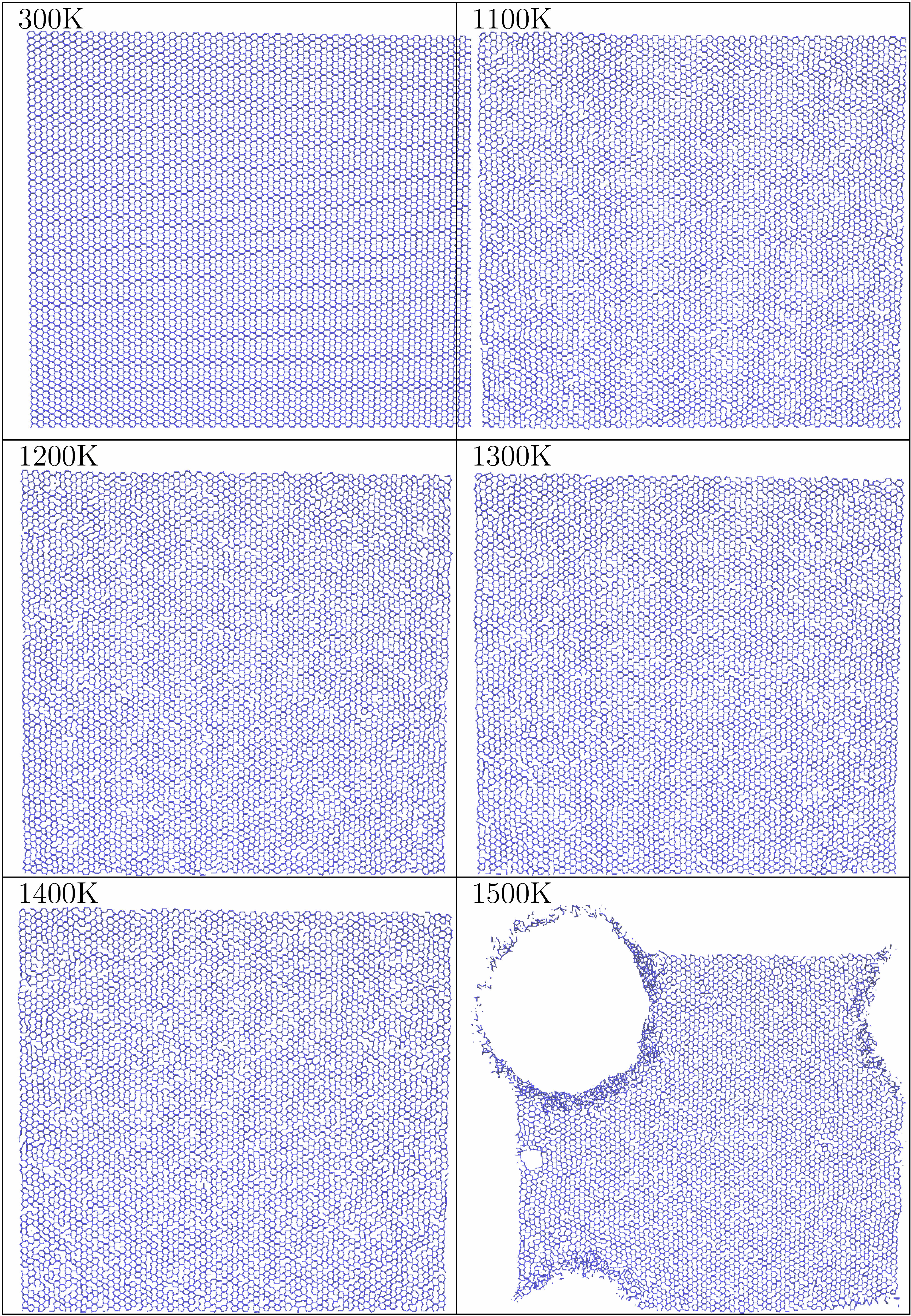}
\end{center}
\caption{Simulated annealing of silicene by using optimized Stillinger-Weber potential. Initially the silicene has equilibrium bond length of 2.5~{\AA}. An abrupt change can be viewed at 1500~K in which the sheet is tearing apart and the melting process has begun. After the melting point, the Si particles settle down to form four smaller ``islands'' which has equilibrium bond length of 2.8~{\AA}.}
\label{t1}
\end{figure}
\end{center}

\begin{center}\begin{figure}[h]
\begin{center}
\includegraphics[width=9.6cm,height=14.4cm]{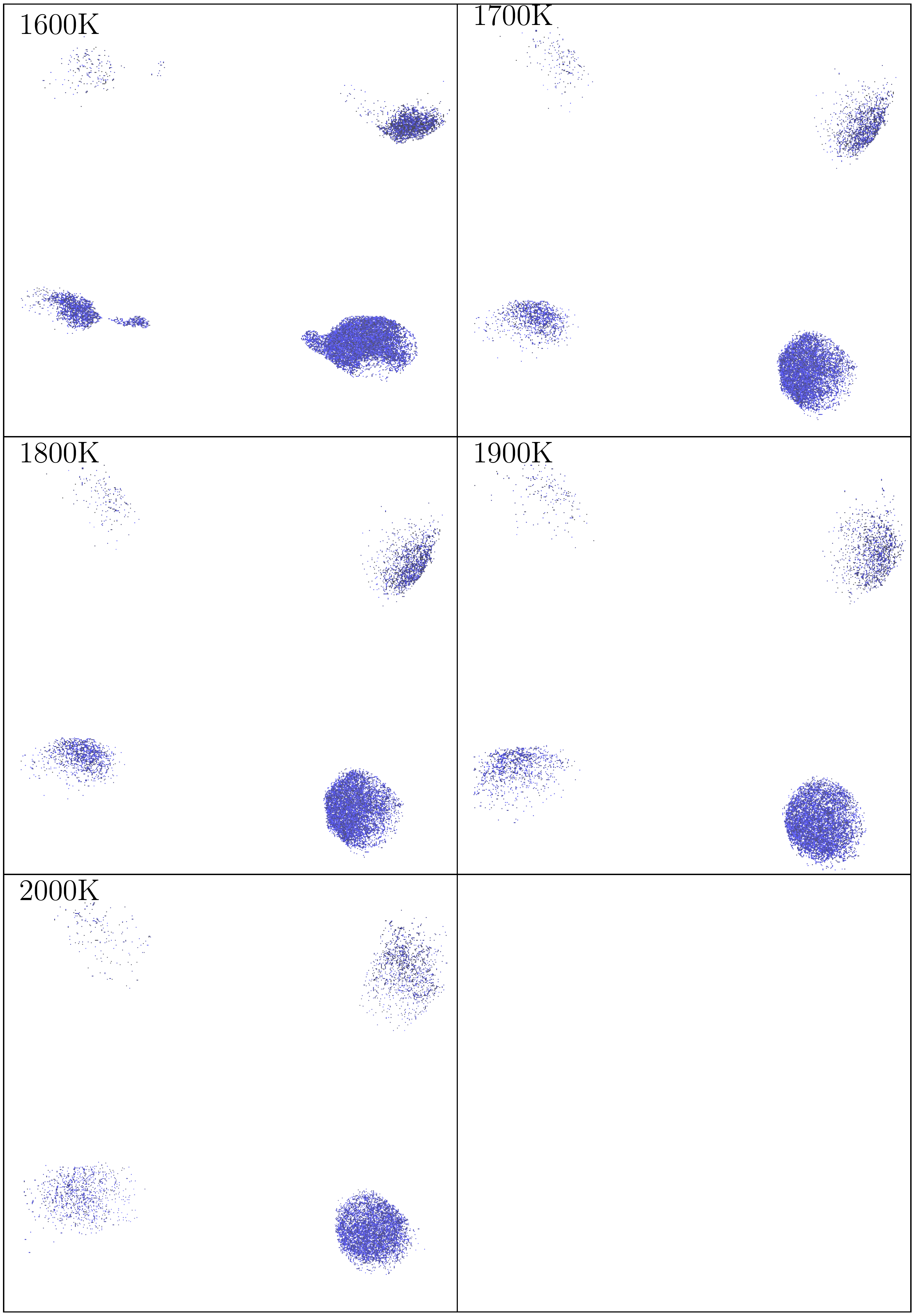}
\end{center}
\caption{Continuation from Fig.~\ref{t1}.}
\label{t2}
\end{figure}
\end{center}

\begin{center}\begin{figure}[p]
\begin{center}
\includegraphics[width=5.4in,height=3.63in]{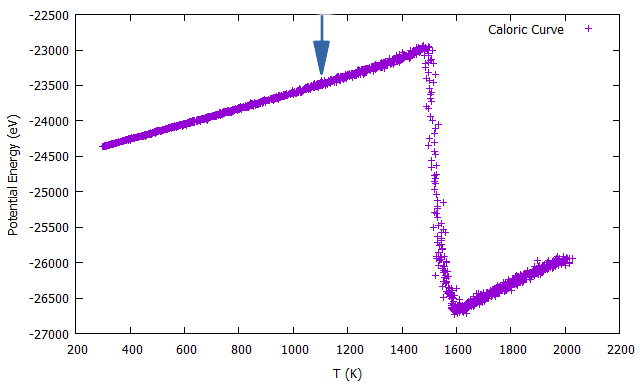}
\end{center}
\caption{Potential energy plot (caloric curve) of the system (eV) against temperature (K) measured for target temperature $T=2000$~K. The arrow indicates the point at which particles begin to vibrate more violently and vigorously such that vacancies begin to appear at selected areas on the silicene sheet as Si particles are ejected from their existing sites. The sharp and abrupt drop of potential energy at 1500~K indicates occurrence of melting at that temperature.}
\label{fig2}
\end{figure}
\end{center}

\begin{center}\begin{figure}[h]
\begin{center}
\includegraphics[width=6.5in,height=3.09in]{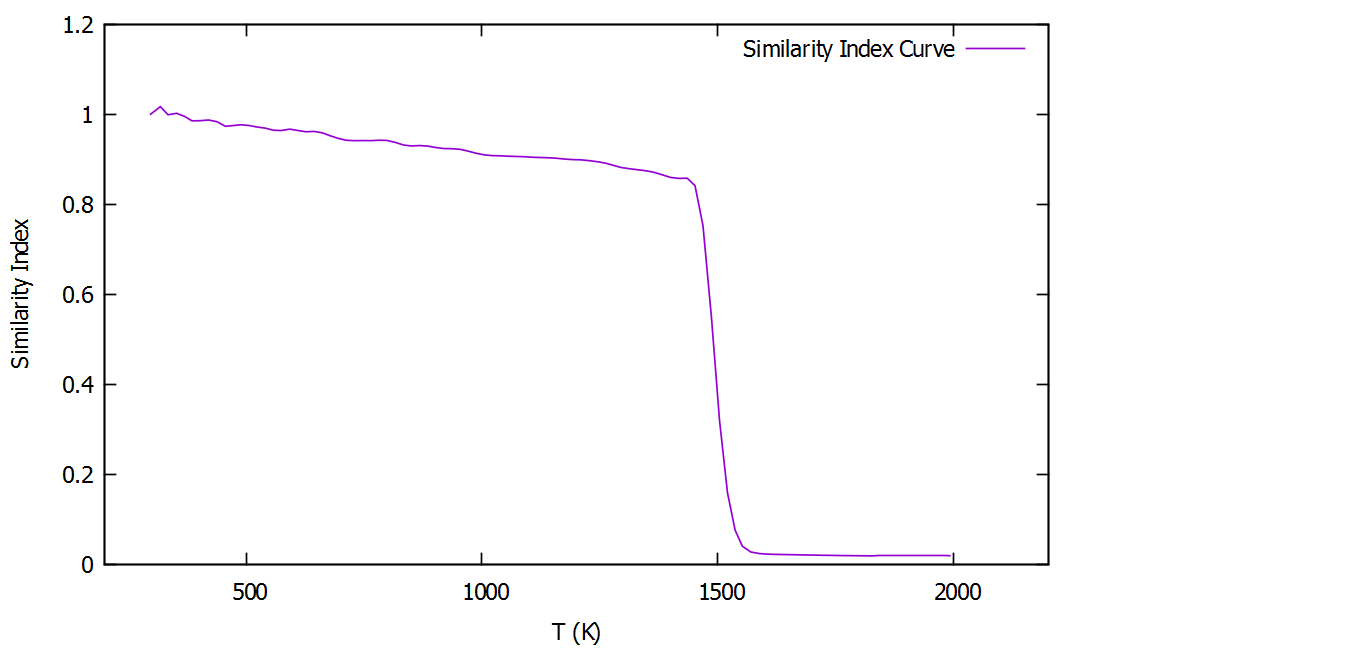}
\end{center}
\caption{Global similarity index plot against temperature. The silicene structure is compared with is 300 K state. The value of 1 means both comparing structure are identical. As the temperature increases, the comparing structures become more and more dissimilar as the kinetic energy of system increases, the Si particles vibrate (there maybe rotational or translational motion based on statistical influence of choice) more violently. The melting occurs at 1500 K, and the structures is immediately unable to identify as a silicene as compare to its 300 K structure.}
\label{figGSI}
\end{figure}
\end{center}

\begin{center}\begin{figure}[h]
\begin{center}
\includegraphics[width=3.44in,height=2.59in]{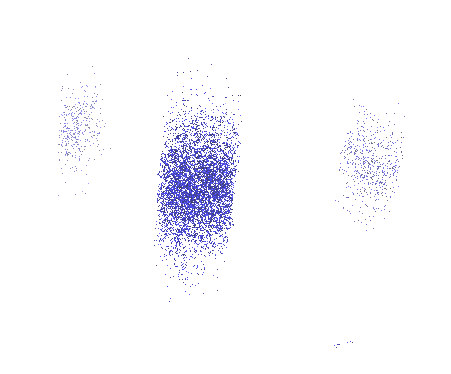}
\end{center}
\caption{Amorphous silicon ``cylinder'' after quenching of the melted silicene. The structure is unable to revert to a sheet form.}
\label{fig5}
\end{figure}
\end{center}

\begin{center}\begin{figure}[h]
\begin{center}
\includegraphics[width=3.44in,height=2.59in]{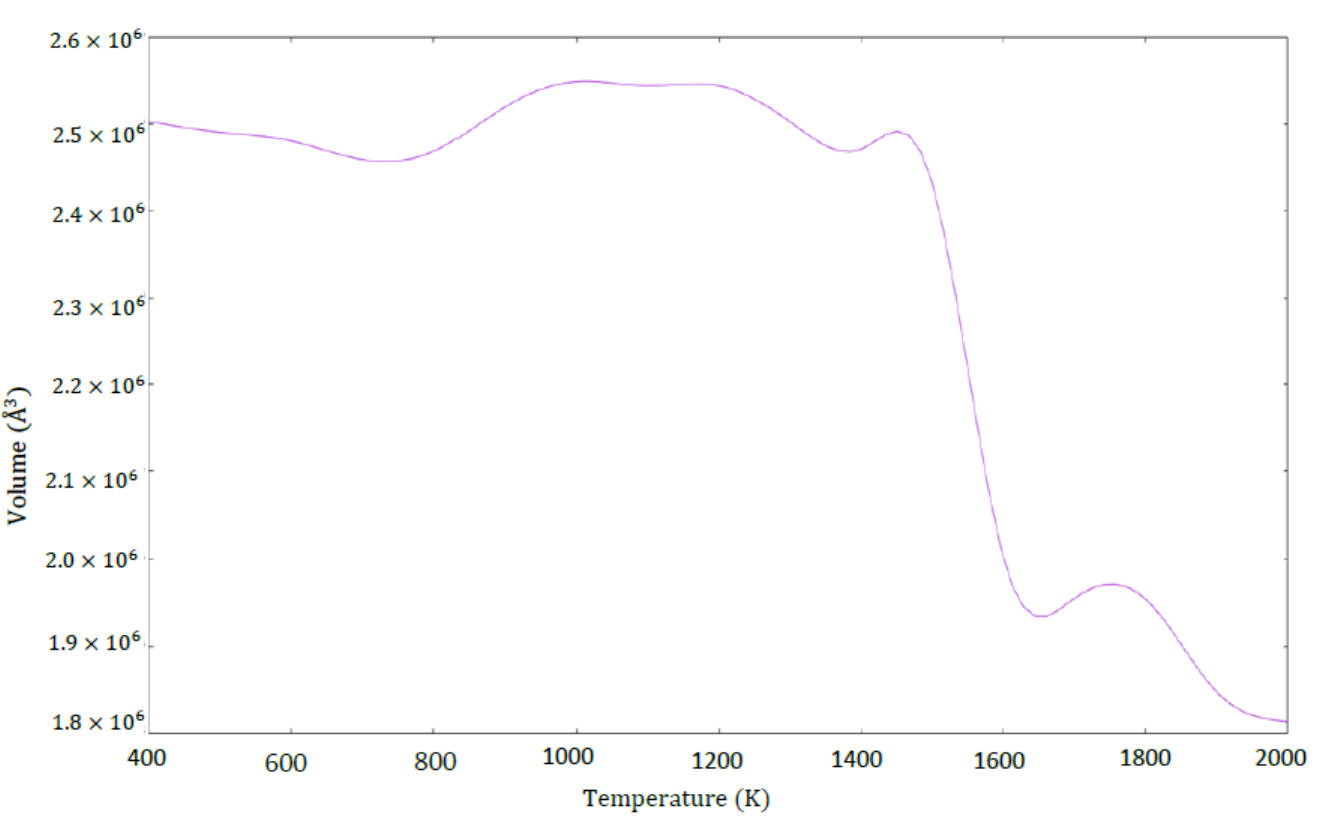}
\end{center}
\caption{Variation of simulation volume as a function of temperature in the NPT ensemble during the melting of the silicene sheet.}
\label{S1}
\end{figure}
\end{center}


\end{document}